\documentclass{amsart}
\usepackage{amsmath, amssymb} 
\usepackage{txfonts}
\newtheorem{Theorem}{Theorem}[section]

\theoremstyle{definition}
\newtheorem{Definition}[Theorem]{Definition}
\theoremstyle{remark}

\numberwithin{equation}{section}

\newcommand{\R}{\mathbb R}
\newcommand{\N}{\mathbb N}

\newcommand{\C}{\mathcal{C}}

\begin{document}

\title{Infinite dimensional integrals beyond Monte Carlo methods: yet another approach to normalized infinite dimensional integrals}
\author{Jean-Pierre Magnot}

\address{Lyc\'ee Blaise Pascal, Avenue Carnot, F-63000 Clermont-Ferrand}
\email{jean-pierr.magnot@ac-clermont.fr}

\maketitle

\begin{abstract} An approach to (normalized) infinite dimensional integrals, including normalized oscillatory integrals, through a sequence of evaluations in the spirit of the Monte Carlo method for probability measures is proposed. in this approach the normalization through the partition function is included in the definition. For suitable sequences of evaluations, the ("classical") expectation values of cylinder functions are recovered \end{abstract}

\section{Introduction} 
Let us first recall two well-developped frameworks.

$\bullet$ Let $(X,\mu)$ be a measured space. Following \cite{Gro}, \cite{Pa}, \cite{Pes}, let us fix a vector subspace $\mathcal{F}\subset L^\infty(X,\mu)$ such that $1_X \in \mathcal{F}.$ A \textbf{mean} on $\mathcal{F}$ is a linear map $\phi:\mathcal{F}\rightarrow \mathbb{C}$
such that $\phi(1_X)=1.$ Alternately, if $(X,d)$ is a metric space, 
given $\mathcal{F}\subset C^0_b(X)$ (space of bounded maps), 
a \textbf{mean} on $\mathcal{F}$ is a linear map 
$\phi:\mathcal{F}\rightarrow \mathbb{C}$ such that $\phi(1_X)=1.$
These two terminologies come from the basic example where $\mu$ is a Borel
probability measure on a metric space $(X,d),$ for which the mean of a continuous integrable map $f$ is its expectation value 
and can be approximated by sequences of barycenters of Dirac measures via Monte Carlo method, namely, via some sequences $(x_n)_{n \in \N} \in X^\N$
such that 
$$ \forall f \in L^1(X,\mu), \int_X f d\mu= \lim_{n \rightarrow +\infty} \frac{1}{n+1} \sum_{k = 0}^n f(x_k).$$
The technical condition for such a sequence is the following: for each $\mu$-measurable set A, 
$\mu(A)= \lim_{n\rightarrow +\infty}\frac{1}{n+1} \sum_{k = 0}^n 1_A(x_k).$

$\bullet$ 
the Feynmann-Kac's formula: It is heuristic in the original Feynman's work, 
and a very difficult question is to 
give it a mathematical (rigorous) sense. Many approaches have been developedin which the following heuristic integral is central: 
$$ \frac{1}{\int e^{-iS}d\lambda} \int f e^{-iS} d\lambda $$
where 
 $S$ is the action functional of the physical theory;
$f$ is $\mathbb{C} -$valued prescribed map defined on an 
infinite dimensional vector  space of configurations; 
$\lambda$ is a heuristic infinite dimensional Lebesgue measure, that is a translation invariant measure on the space of configurations;
$\int e^{-iS}d\lambda$ is a so-called ``normalization constant'',
called ``partition function'' and sometimes noted as $Z(S),$ 
which can be understood as the total volume of the heuristic measure 
$e^{-iS}\lambda,$ of ``density'' $e^{-iS}$ with respect to $\lambda;$ 
and the whole formula stands as a mean value of $f,$ called ``expectation value'' because taken with respect to the (heuristic) probability measure $\frac{1}{\int e^{-iS}d\lambda} e^{-iS} \lambda. $

This too short exposition on Feynman-Kac formula 
is not, of course completely satisfying, compared to
the huge litterature on it and the wide variety of 
tentatives of rigorous definitions.  
We focuse here on the theory of Fresnel integrals, 
which are rigorously defined. The reader can refer to \cite{AB},\cite{AHM}, \cite{AMaz1}, \cite{AMaz2}, \cite{AMaz3}, \cite{Dui}, \cite{ET}, \cite{Ste}.

\vskip 12pt
The technical features of the finite measure setting are almost the same as the
ones of compact spaces. This suggests that, getting into the setting of infinite dimensional integrals, such as heuristic integrals of the Feynman-Kac formula, 
where straight way computations of the normalization constant (the partition function integral) lead to divergent approximations that need to be renormalized (see e.g. \cite{AHM}),
one can expect the same kind of problems in generalizing e.g. a Daniell integral to more complex theories as the problems that occur while passing from compact operators to bounded (or even unbounded) operators acting on a Hilbert space.  
With \cite{Ma10}, we began a research program where 
the normalized integrals with respect to infinite volume measures 
are seen as particular means (which are means spanned by finite measures), 
and are not measures in the strictly speaking sense. 
This approach is coherent with the standard definition of Fresnel integrals  (see e.g. \cite{AHM}) which are defined through sequences of complex 
measures and normalizing weights. The goal of this communication is 
to define a similar appraoach with complex measures 
spanned by Dirac measures, and to compare with 
the classical approach of integration
via cylinder functions.  
\section{Sequences on a separable space and Dirac means}
Let $X$ be a separable topological separable space. 
Given $x \in X$, the evaluation map 
$ \delta_x : f \in \mathbb{C}^X \mapsto f(x) $ is viewed as 
a Dirac measure in a probabilistic way. 
These measures are the extremals of the convex set of Borel probability measures
$P(X).$ Let us define the affine space spanned by these elements. 
  
\begin{Definition}
A complex (resp. real) \textbf{Dirac mean} is a linear map $\tau:\mathcal{D}_\tau \subset C^0_b(X)\rightarrow \C $ which is defined as the limit of barycenters with complex  (resp. real) weights of a sequence of Dirac measures on $X,$ i.e.
for $\mathbb{K}=\mathbb{R}$ or $\mathbb{C},$
$$\exists (x_n, \alpha_n)_{n \in \N}\in (X \times \mathbb{K})^\N, \forall m\in \N^*, $$ $$\left\{ \sum_{n = 0}^{m} \alpha_n \neq 0 \right\} \quad \wedge \quad \left\{ \forall f \in C^0_b(X), \tau(f) = \lim_{m \rightarrow +\infty} \frac{1}{\sum_{n = 0}^{m} \alpha_n} \left( \sum_{n = 0}^{m} \alpha_n \delta_{x_n}(f) \right)\right\}.$$
\end{Definition}
We note by $\widetilde{\mathcal{DM}}_\mathbb{K}(X)$ the space of $\mathbb{K}-$Dirac means, by ${\mathcal{DM}}_\mathbb{K}(X)$ the set of Dirac means $\tau$ such that $\mathcal{D}_\tau = C^0_b(X),$ by $\widetilde{\mathcal{DM}}_\mathbb{R}^+(X)$ the means $\tau$ obtained by a sequence $(\alpha_n)_{n \in \N}\in \R_+^*$ and by ${\mathcal{DM}}_\mathbb{R}^+(X)$ the space $\widetilde{\mathcal{DM}}_\R(X)\cap\widetilde{\mathcal{DM}}_\mathbb{R}^+(X). $

\vskip 12pt
\noindent
\textbf{Examples.}
\begin{itemize}\item \textit{Linear extensions of the limit :}
Let $(x_n)_{n\in \N} \in X^\N$ such that $\lim_{n \rightarrow +\infty}x_n = x.$
Let $(\alpha_n)_{n \in \N} \in \left(\mathbb{R}_+^*\right)^\N$ such that 
$\liminf_{n\rightarrow +\infty} \alpha_n > 0.$ Then, applying the results of \cite{Ma10}, we get that the corresponding Dirac mean $\tau$ equals to the limit at $x.$ By its definition, $\tau$ is defined on a wider class of functions, and then defines a linear extension of the limit. 
\item \textit{Monte Carlo method extended :}
Let $\mu\in P(X)$ and let $(x_n)_{n\in \N}$ such that
 $ \forall f \in L^1(X,\mu), \int_X f d\mu= 
\lim_{n \rightarrow +\infty} \frac{1}{n+1} \sum_{k = 0}^n f(x_k).$
Then for any finite measure $\mu '= \varphi.\mu,$ where $\varphi \in L^\infty$ is the $\mu$-density of $\mu ',$ we have 
$$ \frac{1}{\mu '(X)} \int_X f d\mu'= 
\lim_{n \rightarrow +\infty} \frac{1}{\sum_{k = 0}^n\varphi(x_k)}
 \sum_{k = 0}^n\varphi(x_k)f(x_k).$$ 
If $\mu '$ is only $\sigma$-finite, the density $\varphi$ is 
no longer bounded but the same formula defines $\tau \in \mathcal{DM}_\mathbb{R}(X)$ by 
 $$ \tau(f)= 
\lim_{n \rightarrow +\infty} \frac{1}{\sum_{k = 0}^n\varphi(x_k)}
 \sum_{k = 0}^n\varphi(x_k)f(x_k).$$
For example, if $X = [0;1]$ equipped with the Lebesgue measure $\lambda$ and $\Psi: X \rightarrow \mathbb{R}$ be a smooth diffeomorphismsuch that $\Psi ' >1.$ Let $\mu ' = \Psi^*(\lambda).$ Then, the density $\varphi$ of $\mu '$ with respect to $\lambda$ on $[0;1]$ is bounded below by 1, and hence the corresponding mean $\tau$ is a linear extension of the limit at infinity \cite{Ma10}.
\item \textit{Means on a Hilbert space and on its unit sphere: } A Hilbert space has dense sequences $(x_n)_{n\in \N}$ which are mimicking, by their topological properties, the density property of the sequences used in the classical Monte-Carlo method. Then with the same formula as before, one can take: 
 $$\tau(f) = \lim_{m \rightarrow +\infty} \frac{1}{\sum_{n = 0}^{m} \alpha_n} \left( \sum_{n = 0}^{m} \alpha_n \delta_{x_n}(f) \right)$$
for an adequate function f. We shall discuss a desired class of sequences $(\alpha_n)_{n \in \N}$ next section. But what we have to remark is the following: the unit sphere $S$ of the Hilbert space has the same properies, and then carries also such means. One can even wonder whether radial projections are
possible, and emphasis some invariance under the action of an orthogonal group.
We also remark that many efforts have been made in the context of metric geometry (see e.g. \cite{Gro}, \cite{Pes}) to study means on the sphere. We leave open the question of such a spherical integration, and its significance 
for integration on a ball or on the Hilbert space through   
a spherical-like procedure.

\end{itemize}

 \section{Normalized infinite dimensional integrals as mappings on sequences}
Here we come to the heuristic integral $ \frac{1}{\int e^{-iS}d\lambda} \int f e^{-iS} d\lambda $ on a Hilbert space $H.$ Here, since the part $e^{-iS}$ stands as a density, we can make the following definition:
\begin{Definition} We define a function
\begin{eqnarray*}I : & H^\N \rightarrow & \mathcal{DM}_\C(H) \\
 							& (x_n)_{n \in \N} \mapsto & \tau = \lim_{m \rightarrow +\infty} \frac{1}{\sum_{n = 0}^{m} e^{-iS(x_n)}} \left( \sum_{n = 0}^{m} e^{-iS(x_n)} \delta_{x_n} \right)\end{eqnarray*}
\end{Definition}
Since we have complex coefficients $e^{-iS(x_n)},$ we have to take care about the normalization constant which can be zero for `` bad'' sequences. So that, the domain of $I$ cannot be $ H^\N.$ Then comes the analysis, beyond the convergence of the Monte Carlo method for classical propability measures. Let us quote some open questions:
\begin{itemize}
\item Which kind of sequence lead to invariance by a group acting on $H?$
\item In this general setting, what will be the status of 1-Lipschitz functions that play a crucial role in the approach by metric geometry? 
\item Which sequences lead to (true) measures on $H?$
\item For which perturbations of the action functionnal do we have some
adequate Taylor expansions?
\end{itemize}
The list of open questions can be very long, but we wish to finish by an answer and an open perspective.

\section{Link with the "classical" Monte Carlo method and perspectives}
Let us 
begin with integration of cylinder functions on the infinite cube $[O;1]^\N$ (Daniell integral).  
(heuristic) Wick rotation in order to get a positive measure.
Let us consider now a cylinder function $f.$  Let $P$ be a finite dimensional projection such that $f = f \circ P$ and let $S_P = S\circ P.$  Then, adequate sequences for the Monte  Carlo method are those whose push-forward on $[0;1]^{dim Im P}$  are also adequate for this method. Taking now a creasing sequence of orthogonal projectors
$P_k$ converging (weakly) to identity, the condition on the sequence $(x_n)_{n \in \N}$ is that for each $k \in \N,$ the push-forwards of the sequences $(P_k(x_n))_{n\in \N}$ on $[0;1]^{dim Im P_k}$ fit with the desired conditions. A sequence $(x_n)_{n \in \N}$ for such a method exists, through e.g. the powers of $\piup.$   
Taking an action functionnal $S$ and condidering a finite product measure 
$e^{-S}$ (after Wick rotation), the classical approach of the (classical) Monte-Carlo method is to pull-back the sequences on $[0;1]^\N$ to $\R^\N$ through the coordinatewise pull-back of a product measure. This approach carries no additive weight $(\alpha_n)_{n\in \N}.$ For Fresnel integrals, the density $e^{-iS}$ is approximated by a density $ \xi e^{-iS}$ where the unction $\xi$ is chosen to get two convergent integrals $\int \xi e^{-iS}f d\lambda$ and $\int  \xi e^{-iS} d \lambda.$ Assuming $\xi$ integrable by itself and defined as a product function, there is also (after normalizeation) a possible pull-back of a sequence $(x_n)_{n \in \N}$ adapted for the Monte Carlo method, and the weight we get is only $\alpha_n = e^{-iS(x_n)}.$ 
We recover here an old problem, already quoted in \cite{Ma10}: the ``Lebesgue'' measures on $\mathbb{R}^\N $ have been extensively studied in the 40's but contain very few finite measure subsets (see e.g. \cite{Bak} for an up-to-date exposition). 
As for the example described in \cite{Ma10}, 
the theory of means is an attractive
candidate to complete the theory of probabilities
in such infinite dimensional problems. Unfortunately for applications, 
analysis on such objects has to be developped, and the topolgies of the space $\mathcal{DM}$ have to be studied.

\end{document}